\documentclass[doublecol]{epl2} 

\title{Exceptional points and Bloch oscillations in non-Hermitian lattices with unidirectional hopping}
\shorttitle{Transparency at the interface ... } 

\author{S. Longhi \inst{1,2}}
\shortauthor{S. Longhi \etal}

\institute{                    
  \inst{1}  Dipartimento di Fisica, Politecnico di Milano, Piazza L. da Vinci 32, I-20133 Milano, Italy\\
  \inst{2}  Istituto di Fotonica e Nanotecnlogie del Consiglio Nazionale delle Ricerche, sezione di Milano, Piazza L. da Vinci 32, I-20133 Milano, Italy}

\abstract{The spectral and transport properties of a non-Hermitian  tight-binding lattice with unidirectional hopping are theoretically investigated in three different geometrical settings. It is shown that, while for the infinitely-extended (open) and for the ring lattice geometries the spectrum is complex, lattice truncation makes the spectrum real. However, an exceptional point of order equal to the number of lattice sites emerges. When a homogeneous dc force is applied to the lattice, in all cases an equally-spaced real Wannier-Stark ladder spectrum is obtained, corresponding to periodic oscillatory dynamics in real space. Possible physical realizations of non-Hermitian lattices with unidirectional hopping are briefly discussed.}
\begin{document}

\maketitle

\section{Introduction}
Effective physical models described by non-Hermitian (NH) Hamiltonians are encountered in a wide class of quantum and classical systems  \cite{Moiseyev}.  Examples include open systems and dissipative phenomena in quantum mechanics \cite{Moiseyev,Rotter1,Rotter2,vi}, matter wave systems \cite{ob1,ob2,berry,kor} and optical structures  \cite{Siegman,Makris_PRL_08}.  The physics of NH systems has seen a renewed and growing interest in the past two decades, especially after the introduction of parity-time ($\mathcal{PT}$) symmetric models \cite{Bender_PRL_98,Bender_RPP_2007,Mos1} and their experimental realizations in optical systems \cite{ka1,ka2,ka3}.  Remarkably, in many cases a NH Hamiltonian possess an entire real energy spectrum, in spite of non-self-adjointness. However,  the reality of the spectrum does not ensure
diagonalizability of the Hamiltonian $\mathcal{H}$, which may be prevented by the appearance
of exceptional points (EPs) in the point spectrum of $\mathcal{H}$ \cite{EP0,EP1,EP2}
or spectral singularities in the continuous spectrum of $\mathcal{H}$ \cite{SS,SSbis}. The physical implications of both EPs and spectral singularities have 
been investigated by several authors  (see, for instance,  \cite{EP1,EP2,caz1,palma1,palma2,scappa1,scappa1bis,scappa2,scappa3} and references therein).
Exceptional points correspond to degeneracies
 where both eigenvalues and eigenvectors of a finite-dimensional non-Hermitian Hamiltonian $\mathcal{H}$ coalesce as
a system parameter is varied \cite{EP0,EP1}.  They exhibit non-trivial characteristics compared with those of
most common Hermitian degeneracies, especially concerning adiabatic features and geometric
phases. Among NH Hamiltonian models, 
NH tight-binding lattice networks have become a focal point of research\cite{vi,tb1,tb2,tb3,tb4,tb5,tb7,tb8,tb9,tb10} owing to their versatility, availability of exact
solutions, and possible experimental implementations \cite{Sza}. In a tight-binding lattice, non-Hermiticity is introduced by complex on-site potentials, or by complex or unequal hopping rates (see, for instance, \cite{tb10}).\par 
In this Letter we introduce an exactly-solvable NH tight-binding lattice with {\it unidirectional} hopping, which displays rich physical features. In particular, it is shown that, while the infinitely-extended or ring lattices have a complex energy spectrum, lattice truncation yields an entire real spectrum, however it is highly defective because it is composed by only one energy eigenvalue which is an EP of order equal to the number of lattice sites. We also investigate the transport properties when a homogeneous dc force is applied to the lattice. In this case, the force makes the energy spectrum real  (in the former case) and removes NH degeneracy (in the latter case). An equally-spaced real Wannier-Stark (WS) ladder is obtained in all cases, even for the truncated lattice.  The WS ladder is associated to a periodic dynamics, i.e. Bloch oscillations (BOs), in real space. Possible physical schemes that realize an effective NH lattice with unidirectional hopping are finally discussed. 

\section{Non-Hermitian lattices with unidirectional hopping} 
Let us consider a tight-binding lattice described by the following Hamiltonian
\begin{equation}
\mathcal{H}= \sum_{n} \left( \kappa_1 |n \rangle \langle n+1| + \kappa_2 |n+1 \rangle \langle n | \right)
\end{equation}
where $\kappa_1$ and $\kappa_2$ are the right and left hopping amplitudes, $|n \rangle$ is the Wannier state localized at site $n$. Three different configurations will be considered (see Fig.1): the infinitely-extended (open) lattice, the linear chain comprising $(N+1)$ sites, and the ring lattice comprising $(N+1)$ sites. The Hamiltonian $\mathcal{H}$ is NH whenever $\kappa_2 \neq \kappa_1^*$. Previous works have investigated the cases of unequal hopping ($\kappa_1$ and $\kappa_2$ real but $\kappa_1 \neq \kappa_2$) or dimers with complex hopping ($\kappa_1=\kappa_2$ with non-vanishing imaginary part); see, for instance, \cite{vi,tb7,tb9}. Here we consider the case of {\it unidirectional} hopping by assuming $\kappa_2=0$ and $\kappa_1 \neq 0$. Possible physical implementations of a lattice with unidirectional hopping will be discussed in the last part of the work. If the state vector $| \psi(t) \rangle$ of the system is expanded in series of the Wannier basis $|n \rangle$, $|\psi(t) \rangle=\sum_n c_n(t) |n \rangle$, the evolution of the amplitude probabilities $c_n(t)$ is governed by the following coupled equations  (with $\hbar=1$)
\begin{equation}
i \frac{dc_n}{dt}= \kappa_1 c_{n+1} \; \; \; (n=0, \pm1 , \pm 2 , \pm 3, ....)
\end{equation}
for the infinite lattice [Fig.1(a)], and
\begin{eqnarray}
i \frac{dc_n}{dt} & = &  \kappa_1 c_{n+1} \; \; \; (n=0,1,2,...,N-1) \\
i \frac{dc_N}{dt} & = & 0 
\end{eqnarray}
for the truncated lattice [Fig.1(b)]. For the ring lattice [Fig.1(c)], Eq.(2) holds with the additional cyclic boundary condition $c_{n+N+1}(t)=c_n(t)$. \par
{\it Infinitely-extended lattice}. The eigenstates of $\mathcal{H}$ are Bloch waves of the form $c_n(t)= \exp[iqn-iE(q)t]$, where $ -\pi \leq q < \pi$ is the Bloch wave number and 
\begin{equation}
E(q)=\kappa_1 \exp(iq)= \kappa_1 \cos q+i \kappa_1 \sin q
\end{equation}
 is the energy dispersion curve. Note that the spectrum is absolutely continuous and  the energy has a non-vanishing imaginary part. Since the Bloch states form a complete set of improper functions, the most general solution to the Schr\"{o}dinger equation $i \partial_t | \psi \rangle= \mathcal{H} | \psi \rangle$ is given by
\begin{equation}
c_n(t)= \int_{-\pi}^{\pi}dq F(q) \exp[iqn-iE(q)t]
\end{equation}
where the spectrum $F(q)$ is determined by the initial condition $c_n(0)$, namely $F(q)=(1/ 2 \pi) \sum_n c_n(0) \exp(-iqn)$. Hence $c_n(t)=\sum_l \mathcal{U}_{n,l}(t) c_l(0)$ with $\mathcal{U}_{n,l}(t)=(1/ 2 \pi) \int_{-\pi}^{\pi}dq \exp[iq(n-l)-iE(q)t]$. For example, for single-site excitation, $c_n(0)=\delta_{n,0}$, one has $F(q)=1/(2 \pi)$, and from Eq.(6) after integration one readily obtains
\begin{equation}
c_n(t)=0 \;\; (n >0) \; ,\;\; c_{-n}(t)=\frac{(-i \kappa_1 t)^{n}}{n!} \; \; (n \geq 0).
\end{equation}
{\it Truncated lattice}. For the truncated lattice, from Eqs.(3-4) it readily follows that the energy spectrum is composed solely by the value $E=0$, with corresponding eigenvector $(c_0,c_1,c_2,...,c_N)^T=(1,0,0,0,...,0)^T$. Interestingly, the energy $E=0$ is an EP of order $(N+1)$. This follows from the fact that the $(N+1) \times (N+1)$ matrix $\mathcal{H}_{n,m}= \langle n | \mathcal{H} m \rangle$ of the linear system of equations (3-4) is in the Jordan canonical form
\begin{equation}
\mathcal{H}_{n,m}= \left( 
\begin{array}{ccccccc}
0 & \kappa_1 & 0 & 0 & ...& 0  & 0  \\
0 & 0 & \kappa_1 & 0 & ... & 0  & 0 \\
... & ... & ... & ... & ... & ... & ... \\

0 & 0 &  0& 0 & ... & 0 & \kappa_1 \\
0 & 0 &  0& 0 & ... & 0 & 0 
\end{array}
\right).
\end{equation}
Hence, lattice truncation makes the spectrum of $\mathcal{H}$ entirely real, however it introduces an exceptional point of  order $(N+1)$. Since $E=0$ is a defective eigenvalue, associated eigenfunctions should be included to obtain a complete basis. The most general solution to Eqs.(3,4) is readily obtained by direct integration. For example, the solution corresponding to single site excitation, $c_n(0)=\delta_{n,n0}$ with $0 \leq n_0 \leq N$, is given by
\begin{equation}
c_n(t) = \left\{ 
\begin{array}{cc}
 0 & n=N,N-1,..,n_0+1 \\ 
 \frac{(-i \kappa_1 t )^{n_0-n}}{(n_0-n)!}  & n=0,1,...,n_0
 \end{array}
 \right.
\end{equation}
Note that, even though the spectrum of $\mathcal{H}$ is real, a secular growth of the amplitudes $c_n$ is found like in the infinitely-extended lattice. The reason thereof is that $E=0$ is an EP.\\
{\it Ring lattice}.  For the ring lattice of Fig.1(c), the eigenstates of $\mathcal{H}$ are Bloch waves like for the infinitely-extended lattice, namely $c_n(t)=\exp[iqn-iE(q)t]$ where $E(q)=\kappa_1 \exp(iq)$ and the Bloch wave number $q$ is quantized owing to the cyclic period boundary conditions, i.e. $q=q_k=2 \pi k /(N+1)$ ($k=0,1,2,...,N$). The energy spectrum is thus point-like and complex, comprising $(N+1)$ distinct complex energies. Note that, as opposed to the finite lattice chain of Fig.1(b), in the ring geometry the lattice Hamiltonian $\mathcal{H}_{n,m}$ is not in the Jordan canonical form like in Eq.(8), because $\mathcal{H}_{n,m}$ differs from the canonical form (8) for the non-vanishing element
$\mathcal{H}_{N,0}=\kappa_1$ that arises from the additional coupling of lattice sites $n=N$ and $n=0$ in the ring geometry. This makes the $(N+1)$ eigenvalues of $\mathcal{H}_{n,m}$ distinct and hence there is not an EP in the ring geometry.
The most general solution to Eq.(2) can be calculated following the same procedure like  for the infinitely-extended lattice, provided that the Boch wave number $q$ is quantized. One obtains $c_n(t)=\sum_{l=0}^{N} \mathcal{U}_{n,l}(t)c_l(0)$ with $\mathcal{U}_{n,l}(t)=(N+1)^{-1} \sum_{q_k} \exp[iq_k(n-l)-iE(q_k)t]$. Note that, as $N \rightarrow \infty$, the sum over $q_k=2 k \pi/(N+1)$ in the previous expression of $\mathcal{U}_{n,l}(t)$ can be replaced by an integral over $q$, obtaining the limiting case of the infinitely-extended lattice discussed above. 
\begin{figure}
\onefigure[width=8.8cm]{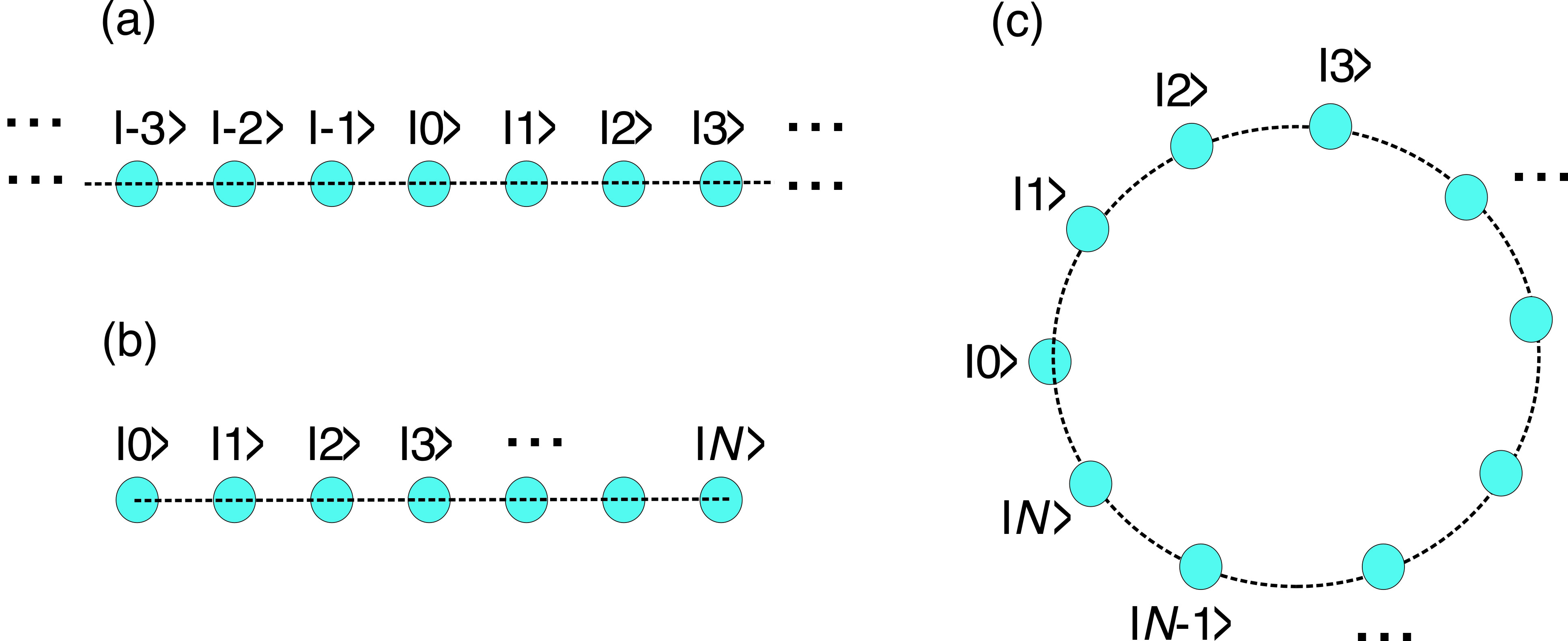}
\caption{(Color online) Schematic of a tight-binding lattice in three different geometrical settings: (a) infinitely-extended (open) linear chain, (b) finite linear chain formed by $(N+1)$ sites, and (c) ring lattice composed by $(N+1)$ sites.}
\end{figure}

\section{Wannier-Stark energy spectrum and Bloch oscillations} If a homogeneous dc force $F$ is applied to a linear chain with unidirectional hopping, the Hamiltonian of the system for the infinitely-extended lattice or for the finite lattice chain [Figs.1(a) and (b)] reads
\begin{equation}
\mathcal{H}= \sum_{n}  \kappa_1 |n \rangle \langle n+1| +\sum_n Fn  |n \rangle \langle n | 
\end{equation}
{\it Infinitely-extended lattice.} For the infinitely-extended lattice, the energy spectrum $E$ of $\mathcal{H}$ can be calculated by a standard method. Let $c_n(t)=a_n \exp(-iEt)$ be an eigenfunction of $\mathcal{H}$ with energy $E$, i.e. $Ea_n=\kappa_1 a_{n+1}+Fna_n$, and let us switch from the Wannier basis to the Bloch representation by introduction of the Bloch wave function $S(q)=\sum_{n}a_n \exp(iqn)$ with momentum $q$. It then follows that $S(q)$ satisfies the differential equation
\begin{equation}
i F \frac{dS}{dq}=\left\{ \kappa_1 \exp(-iq)-E \right\}S
\end{equation}
which is readily solved, yielding
\begin{equation}
S(q)=S(0) \exp \left\{ i E q/F+( \kappa_1 /F) \left[ \exp(-iq)-1\right] \right\} 
\end{equation}
The allowed energies $E$ are obtained by imposing the periodic boundary condition for $S(q)$, i.e. $S(2 \pi)=S(0)$. This yields
\begin{equation}
E_l=Fl \; \; (l=0, \pm 1, \pm 2, \pm 3, ...).
\end{equation}
The corresponding eigenfunctions $a_n^{(l)}$ can be then obtained from the inversion relation $a_n= (1 / 2 \pi) \int_{-\pi}^{\pi} dq S(q) \exp(-iqn)$. One obtains
\begin{equation}
a_{n}^{(l)}= \left\{
\begin{array}{cc}
0 & n \geq l+1 \\
1 & n=l \\
\left( \frac{\kappa_1}{F} \right)^{l-n} \frac{1}{(l-n)!}  &  n<l
\end{array}
\right.
\end{equation}
Note that, since $\sum_n |a_{n}^{(l)}|^2 < \infty$, the energies $E_l$ belong to the point spectrum of $\mathcal{H}$ and form an equally-spaced real Wannier-Stark ladder. Hence, application of a dc force to the lattice changes the spectrum from a continuous complex energy band into an equally-spaced real Wannier-Stark ladder. In real space, the equally-spaced Wannier-Stark ladder spectrum yields an oscillatory motion, similar to BOs of ordinary (Hermitian) lattices. The oscillatory motion is not secularly damped nor amplified. Note that previous results on BOs in $\mathcal{PT}$ symmetric optical lattices \cite{LonghiPRL09} generally yield {\it complex} Wanner-Stark ladders, with either amplified or damped BOs depending on the sign of the force. 

{\it Truncated lattice}.
The energy spectrum of $\mathcal{H}$ for the truncated lattice can be readily calculated from the form of the $(N+1) \times (N+1)$ matrix $\mathcal{H}_{n,m}=\langle n| \mathcal{H}  m \rangle$ in the Wannier basis, which reads 
\begin{equation}
\mathcal{H}_{n,m}= \left( 
\begin{array}{ccccccc}
0 & \kappa_1 & 0 & 0 & ...& 0  & 0  \\
0 & F & \kappa_1 & 0 & ... & 0  & 0 \\
0 & 0 & 2 F & \kappa_1 & ... & 0  & 0 \\

... & ... & ... & ... & ... & ... & ... \\

0 & 0 &  0& 0 & ... & F(N-1) & \kappa_1 \\
0 & 0 &  0& 0 & ... & 0 & FN 
\end{array}
\right).
\end{equation}
Since $\mathcal{H}_{n,m}$ is in the canonical Jordan form, the eigenvalues of $\mathcal{H}_{n,m}$ are the elements on the main diagonal, i.e. one has
\begin{equation}
E_l=lF \; \; (l=0,1,2,...,N).
\end{equation}
The corresponding eigenvectors, $c_n^{(l)}(t)=a_{n}^{(l)} \exp(-iE_l t)$, are given by
\begin{equation}
a_{n}^{(l)}= \left\{
\begin{array}{cc}
0 & n \geq l+1 \\
1 & n=l \\
\left( \frac{\kappa_1}{F} \right)^{l-n} \frac{1}{(l-n)!} &0 \leq  n<l
\end{array}
\right.
\end{equation}

\begin{figure}
\onefigure[width=8.8cm]{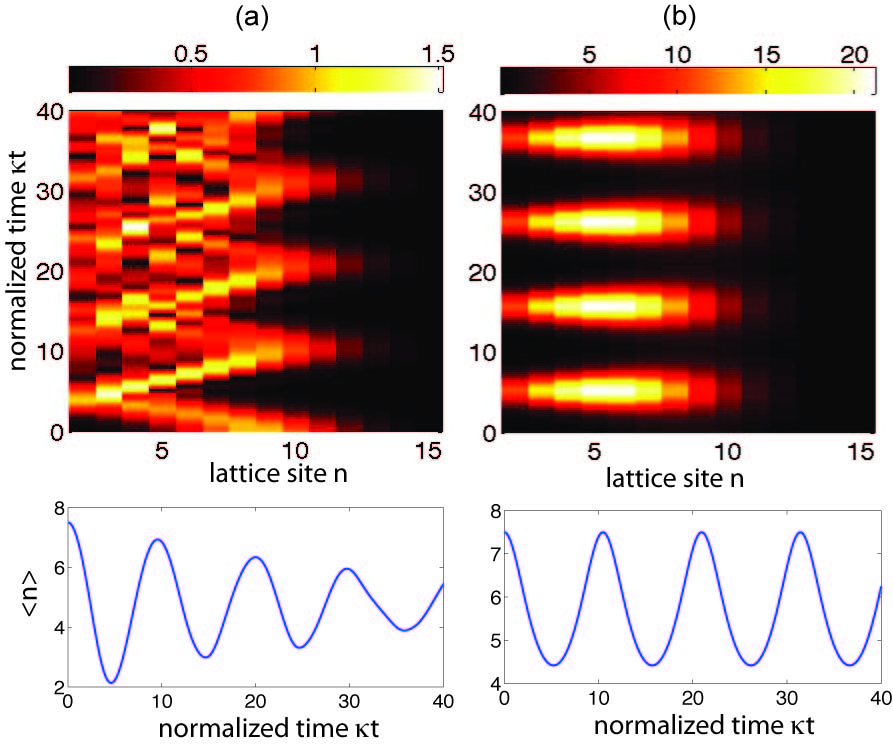}
\caption{(Color online) BOs in a truncated lattice made of $N=15$ sites with an applied dc force $F/ \kappa=-0.6$ for  (a) an Hermitian lattice ($\kappa_1=\kappa_2=\kappa$), and  (b) a NH lattice with unidirectional hopping ($\kappa_1=\kappa$, $\kappa_2=0$). The upper panels show snapshots of $|c_n(t)|^2$, whereas the lower panels show the trajectory $\langle n(t)) \rangle$ of the wavepacket center of mass. The initial condition is $c_n(t)=\exp[-(n-N/2)^2/w^2]$ with $w=3$.}
\end{figure}

Hence in the truncated lattice application of a dc force removes the EP point, the energy spectrum remains entirely real and it is composed by a set of equally-spaced $(N+1)$ energies [Eq.(16)]. Correspondingly, the dynamics in real space is still periodic, in spite of lattice truncation. Note that such a result is quite remarkable and it has no counterpart in Hermitian lattices, where it is well known that lattice truncation smears the periodic oscillatory motion when a wave packet reaches the lattice boundary. To clarify this point, in Fig.2 we show as an example the numerically-computed evolution of  a wave packet (snapshot of $|c_n(t)|^2$) and of its path $\langle n(t) \rangle = \sum_n n |c_n(t)|^2 / \sum_n |c_n(t)|^2$ in a truncated lattice with an applied dc force, for either an Hermitian lattice [$\kappa_1=\kappa_2=\kappa$ real, Fig.2(a)] and a NH lattice with unidirectional hopping [$\kappa_1=\kappa$, $\kappa_2=0$, Fig.2(b)]. Note that in the former case the dynamics is clearly aperiodic and the BOs of the wave packet are smeared out,  whereas in the latter case the dynamics is exactly periodic with period $T_B=2 \pi /F$ and BOs are not degraded.\\
{\it Ring Lattice.} In the ring lattice geometry of Fig.1(c), the effect of a dc force along the ring is generally obtained by application of a magnetic flux $\Phi(t)$ threading the ring that linearly increases in time. Assuming $\Phi(t)=\Phi_0t$ in units of the flux quantum $h/e$, the magnetic flux introduces a Peierls phase $Ft$ in the hopping rate $\kappa_1$ that linearly increases in time with a rate $F$, where $F=2 \pi \Phi_0/(N+1)$ (see, for instance, \cite{Citrin,Orellana}). The amplitude probabilities $c_n$ of particle occupation at the various lattice sites on the ring then satisfy the coupled equations
\begin{equation}
i \frac{dc_n}{dt}=\kappa_1 \exp(iFt) c_{n+1}
\end{equation}
($n=0,1,2,...,N$) with the cyclic periodic boundary condition $c_{N+1}(t)=c_0(t)$. Owing to the periodicity in time of the coefficients entering in Eq.(18), Floquet theory applies and the solutions to Eq.(18) are of the form $c_n(t)=u_n(t) \exp(-i \mu t)$, where $u_n(t+T_B)=u_n(t)$, $T_B=2 \pi /F$ and $\mu$ is the quasi-energy, which is assumed to vary in the range $(-F/2,F/2)$. The $(N+1)$ quasi energies $\mu=\mu_l$ can be readily determined following a similar procedure than that used for the infinitely extended lattice, by switching from the Wannier to the Bloch representations with the introduction of the Bloch-Floquet function $S(q_l,t)=\sum_{n=0}^N u_n(t) \exp(iq_l n)$, where the Bloch wave number $q_l$ is quantized according to $q_l=2 \pi l/(N+1)$ ($l=0,1,2,3,...,N$). For construction, $S(q_l,t)$ is periodic in time with period $T_B$. Using Eq.(18) it readily follows that $S$ satisfies the equation
\begin{equation}
 \frac{dS(q_l,t)}{dt}= i \left[ \mu- \kappa_1 \exp(iFt-iq_l) \right] S
\end{equation}
which is solved by
\begin{eqnarray}
S(q_l,t) & = & S(q_l,0)  \\
& \times & \exp \left[i \mu t- i \kappa_1 \int_0^t dt' \exp(iFt'-iq_l) \right]. \nonumber
\end{eqnarray}
The condition $S(q_l,T_B)=S(q_l,0)$ then yields for the quasi energies $\mu_l$ the values
\begin{equation}
\mu_l=\frac{\kappa_1}{T_B} \int_{0}^{T_B} dt' \exp(iFt'-iq_l)
\end{equation}
$(l=0,1,2,...,N$). Since the integral on the right hand side of Eq.(21) vanishes, it follows that the $(N+1)$ quasi energies  $\mu_l$ collapse to the value $\mu=0$. This shows that the quasi energy spectrum is real and the dynamics is fully periodic with period $T_B= 2 \pi/ F$, like in the previously discussed cases. Such an oscillatory dynamics corresponds to magnetic BOs \cite{Citrin} in a NH ring.

\section{Realization of NH lattices with unidirectional hopping} NH lattices with unequal hopping rates $\kappa_1$ and $\kappa_2=1/ \kappa_1$ were introduced in a seminal paper by Hatano and Nelson using a superconducting structure in a magnetic field \cite{vi}, however such a system does not allow to realize the unidirectional hopping regime because of the divergence of $\kappa_1$ as $\kappa_2 \rightarrow 0$. Here we suggest a different route, which is based on a suitable periodic temporal modulation of {\it complex} site potentials in a lattice with Hermitian hopping. Let us consider a NH lattice with real and symmetric hopping rate $\kappa$ and with site energies $V_n(t)$, which is described by the Hamiltonian
\begin{equation}
\mathcal{H}_1= \sum_{n} \kappa \left( |n \rangle \langle n+1| + |n+1 \rangle \langle n | \right)+\sum_n V_n(t) |n \rangle \langle n|
\end{equation}
We assume that the site potentials $V_n(t)$ are periodically modulated in time with a period $T=2 \pi / \omega$. Such an Hamiltonian can describe, for instance, propagation of light waves in an array of equally-spaced optical waveguides with longitudinally-modulated propagation constants (the real part of $V_n$) and optical amplification or loss (the imaginary part of $V_n$).  In the fast modulation regime $ \omega \gg \kappa$,  application of the rotating-wave approximation yields the effective Hamiltonian with renormalized hopping rates (see, for instance, \cite{Creff,Crespi})
\begin{equation}
\mathcal{H}_{eff} \simeq \sum_n \left( \rho_n |n+1 \rangle \langle n | + \sigma_n |n \rangle \langle n+1| \right)
\end{equation}
where we have set
\begin{eqnarray}
\rho_n & =  & \kappa \langle \exp \left[ i \int_0^t dt' (V_n-V_{n+1} )\right] \rangle \\
\sigma_n & = & \kappa \langle \exp \left[ i \int_0^t dt' (V_n-V_{n-1} )\right] \rangle \
\end{eqnarray}
and where $ \langle ... \rangle = (1/T) \int_0^T dt ...$ denotes the time average over one oscillation cycle. To realize a lattice with unidirectional hopping, the site potentials $V_n(t)$ should be chosen such that $\rho_n=\kappa_1 \neq 0$ and $\sigma_n=0$. This requires necessarily to introduce complex site potentials $V_n(t)$. A possibility is the following one:
\begin{equation}
V_n(t)= 
 \theta n \delta(t-T_1) +\frac{1+(-1)^n}{2}(\alpha+i \beta) H(t)
\end{equation}
for $0^+ <t \leq  T^+$ and with $V_n(t+T)=V_n(t)$. In Eq.(26),  $\theta$, $\alpha$ and $\beta$ are real amplitudes, $0<T_1<T$, whereas the function $H(t)$ is defined by
\begin{eqnarray}
H(t)= \left\{
\begin{array}{cc}
1 & 0<t<T_1/4 \\
-1 & T_1/4 < t < 3T_1/4 \\
1 & 3T_1/4 <t< T_1 \\
0 & T_1<t<T.
\end{array}
\right.
\end{eqnarray}
In an optical waveguide setting, the modulation defined by Eqs.(26) and (27) corresponds to segmented waveguides at even sites, with a first segment of complex propagation constant detuning $(\alpha+i \beta)$  ($0<t<T_1/4$), a second segment with detuning $-(\alpha+i \beta)$  ($T_1/4<t<3T_1/4$), a third segment with detuning $(\alpha+i \beta)$  ($3T_1/4<t<T_1$), and a last  waveguide segment with unmodified index ($T_1<t<T$). A lumped transverse phase gradient  $n \theta$ is also impressed at $t=T_1$; see Fig.3(a) for a schematic. Lumped phase shifts can be introduced using e.g. a zig-zag axis profile or by short waveguide segmentation, as demonstrated in Refs.\cite{fase1,fase2}. 
Substitution of Eqs.(26,27) into Eqs.(24,25) yields
\begin{eqnarray}
\rho_n & = & x \frac{\sin \Gamma}{\Gamma}+(1-x) \exp(-i \theta) \equiv \kappa_1 \\
\sigma_n & = & x \frac{\sin \Gamma}{\Gamma}+(1-x) \exp(i \theta) \equiv \kappa_2
\end{eqnarray}
where we have set
\begin{equation}
x \equiv \frac{T_1}{T} \; , \;\;\; \Gamma \equiv \frac{(\alpha+i\beta)T_1}{4}
\end{equation}
Note that both $\rho_n$ and $\sigma_n$ do not depend on the index $n$. The condition $\sigma_n=\kappa_2=0$ can be realized, for example, by assuming $\theta=\pi/2$, $x=0.8$ and $\Gamma \simeq 3+0.7i$. Correspondingly, one has $\kappa_1 \simeq - 0.4i  \kappa$. For the observation of BOs, a transverse index gradient $Fn$ can be introduced by e.g. circularly-curving the waveguides along the axial propagation distance $t$.\\

\begin{figure}
\onefigure[width=8.8cm]{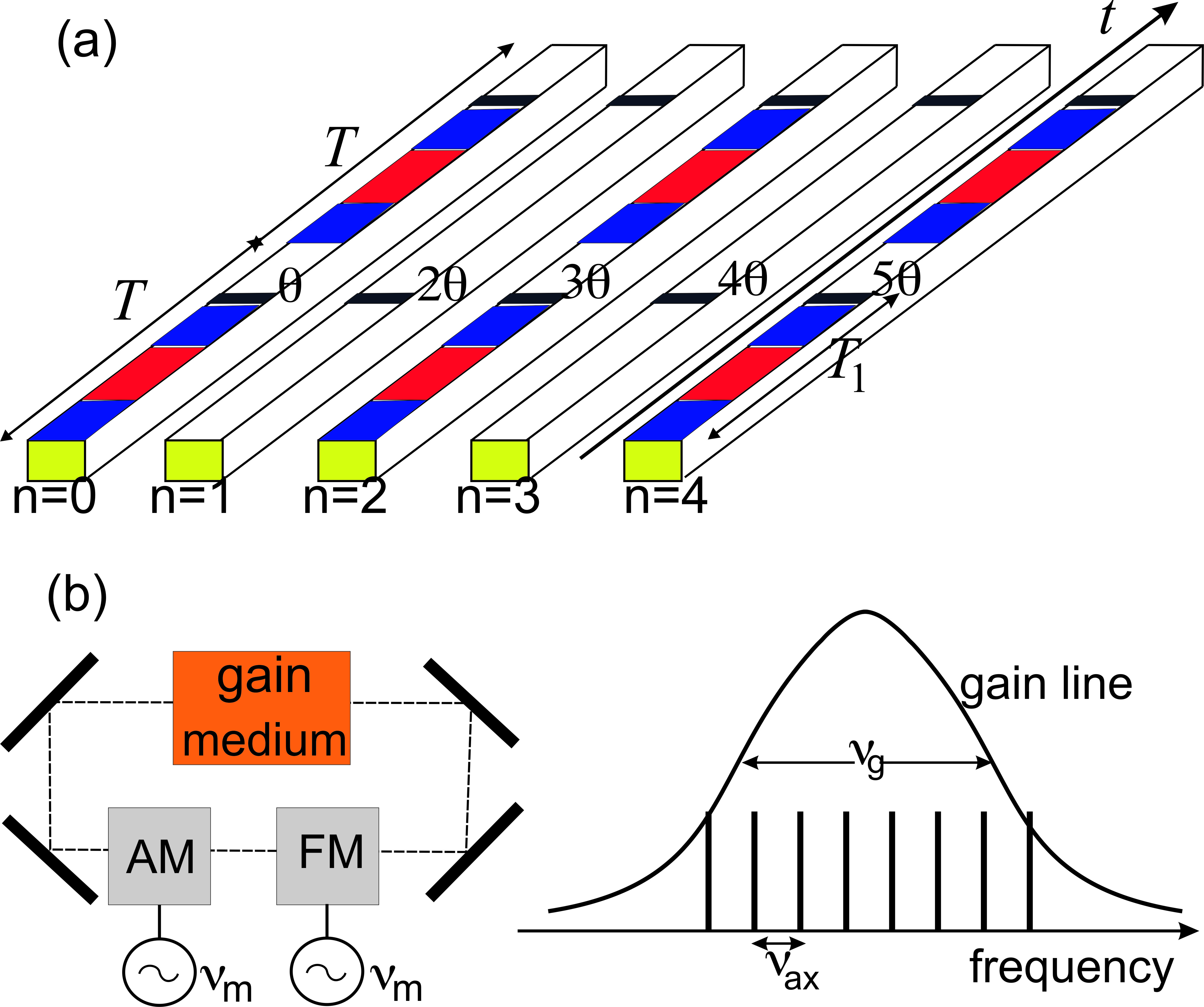}
\caption{(Color online) Possible physical realizations of a NH lattice with unidirectional hopping. (a) Light propagation in an array of evanescently-coupled longitudinally-segmented optical waveguides. Segmentation is periodic with spatial period $T$, occurs in waveguides with even order and corresponds to alternating sections of gain/loss and high/low index (shaded regions of length $T_1$ in the figure). A lumped transverse phase gradient is applied at distances $t=T_1, T_1+T, T_1+2T,...$ to all waveguides. The exact profile of segmentation is defined by Eq.(26) and (27) given in the text. (b) Schematic of a mode-locked laser with amplitude (AM) and phase (FM) modulators sinusoidally driven at a frequency $\nu_m$ close to the axial mode separation $\nu_{ax}$ of the laser cavity. The temporal evolution of of spectral mode amplitudes (the comb of modes in the right panel, spaced by $\nu_{ax}$) emulates an Hermitian lattice when the amplitude modulator is switched off, and a NH lattice with unidirectional hopping when both phase and amplitude modulators are switched on.}
\end{figure}

Finally, it is worth briefly mentioning another physical system where the NH lattice Hamiltonian (1) with unidirectional hopping ($\kappa_2=0$) and a dc force could be realized. This system is provided by light dynamics in a mode-locked laser with amplitude and phase modulators placed inside the laser cavity \cite{Longhi1,Longhi2}; see Fig.3(b) for a schematic. The evolution of the amplitudes $c_n(t)$ of the cavity axial modes in the laser are governed by the following set of coupled equations (see, for instance, \cite{Longhi1} )
\begin{eqnarray}
i\frac{dc_n}{dt} & = & [Fn+i(g-l)-iD_gn^2]c_n+ \Delta_{FM} (c_{n+1}+c_{n-1}) \nonumber \\
& + & i \Delta_{AM}[ \exp(i \varphi) c_{n+1}+\exp(-i \varphi) c_{n-1}] 
\end{eqnarray}
where $g$ and $l$ are the single-pass saturated gain and cavity loss rate, respectively, $t$ measures the round-trip number, $F= 2 \pi (\nu_m-\nu_{ax})/ \nu_{m}$ is a normalized detuning parameter between the modulation frequency $\nu_m$ and the cavity axial mode separation $\nu_{ax} \simeq \nu_m$, $D_g=(\nu_m/ \nu_g)^2$, $\nu_g$ is the line width of the lasing transition, $\Delta_{AM}$, $\Delta_{FM}$ are the modulation depths impressed by the amplitude and phase modulators, respectively, and $\varphi$ is the phase offset between the two modulators. We note that, when the amplitude modulator is switched off  ($\Delta_{AM}=0$) and neglecting finite bandwidth effects of the gain medium (i.e. assuming $D_g \simeq 0$), at the onset of lasing $g =l$ Eqs.(31) describe ordinary BOs of an Hermitian lattice for the spectral modes $c_n$. Such a kind of BOs can be observed as a breathing dynamics of the laser spectrum in the transient switch-off of the laser, as demonstrated in the experiment of Ref.\cite{Longhi1}.  Let us now consider the case where both the amplitude and phase modulators are switched on, and let us assume $\Delta_{AM}=\Delta_{FM}$ and tune the phase $\varphi$ to the value $\varphi=-\pi/2$. In this case, neglecting finite bandwidth effects, at the onset of lasing ($g=l$) the dynamical evolution of the spectral axial amplitudes $c_n$ emulates the NH Hamiltonian (10) with $\kappa_1=2\Delta_{AM}$.
\section{Conclusions} 
In this work we have investigated the spectral and transport properties of an exactly-solvable NH tight-binding lattice with unidirectional hopping in three different configurations: infinitely-extended (open) linear lattice, finite linear chain, and ring lattice (Fig.1). Whereas for the infinitely-extended and ring lattices the energy spectrum is complex, in a finite linear chain  a single and real energy eigenvalue is found, which is an exceptional point of high order. In all geometries application of a dc force yields a real Wannier-Stark ladder, corresponding to exact periodic dynamics in real space (Bloch oscillations). This  occurs even for the truncated (finite chain) lattice. Such a result is a clear signature of unidirectional hopping and it is not found in ordinary Hermitian lattices, where lattice truncation smears out the Bloch oscillations. Two possible physical realizations of NH lattices with unidirectional hopping have been suggested, based on light transport in engineered optical waveguide lattices with segmented regions of alternating gain/loss and index contrast regions, or  on light dynamics in a mode-locked laser with intracavity amplitude and phase modulators.


\end{document}